\documentclass[12pt,preprint]{aastex}
\usepackage{times}

\newcommand{\ms}{m~s$^{-1}$}
\newcommand{\dtm}{$\delta\tau_\mathrm{MF}$}
\newcommand{\dtc}{$\delta\tau_\mathrm{CtoL}$}

\shorttitle{Solar Meridional Circulation and Center-to-Limb Effect}
\shortauthors{Chen \& Zhao}

\begin{document}
\title{A Comprehensive Method to Measure Solar Meridional Circulation
and Center-to-Limb Effect Using Time--Distance Helioseismology}

\author{Ruizhu Chen\altaffilmark{1,2} and Junwei Zhao\altaffilmark{2}}
\altaffiltext{1}{Department of Physics, Stanford University, Stanford, 
CA 94305-4060}
\altaffiltext{2}{W.~W.~Hansen Experimental Physics Laboratory, Stanford 
University, Stanford, CA 94305-4085}

\begin{abstract}
Meridional circulation is a crucial component of the Sun's internal dynamics,
but its inference in the deep interior is complicated by a systematic 
center-to-limb effect in helioseismic measurement techniques. Previously,
an empirical method, removing travel-time shifts measured for east-west traveling
waves in the equatorial area from those measured for north-south traveling waves 
in the central meridian area, was used, but its validity and accuracy need to 
be assessed. Here we develop a new method to separate the center-to-limb 
effect, \dtc, and meridional-flow-induced travel-time shifts, \dtm, in a more 
robust way. Using 7-yr observations of the {\it SDO}/HMI, we
exhaustively measure travel-time shifts between two surface locations along
the solar disk's radial direction for all azimuthal angles and all skip
distances. The measured travel-time shifts are a linear combination of
\dtc\ and \dtm, which can be disentangled through solving the linear
equation set. The \dtc\ is found isotropic relative to the azimuthal angle,
and the \dtm\ are then inverted for the meridional circulation. Our
inversion results show a three-layer flow structure, with equatorward
flow found between about 0.82 and 0.91 R$_\sun$ for low latitude areas
and between about 0.85 and 0.91 R$_\sun$ for higher latitude areas. Poleward
flows are found below and above the equatorward flow zones, indicating a
double-cell circulation in each hemisphere.

\end{abstract}

\keywords{Sun: helioseismology --- Sun: oscillations --- Sun: interior }

\section{Introduction}
\label{sec1}
Solar meridional circulation is crucial to the understanding of the Sun's
interior dynamics and dynamo. It is widely believed that the Sun's
differential rotation and convection drive the solar dynamo \citep{par55},
and the meridional circulation, although one order of magnitude smaller
than the differential rotation, plays an important role in transporting
magnetic flux and redistributing angular momentum \citep[e.g.,][]{Wang89, 
Mie05, Upton14a, Upton14b, Fea15}. The speed of meridional circulation
is also thought to be related to the duration and
amplitude of solar cycles \citep{Hath10, Dik10}. The profile of      
meridional circulation has long been sought, but for many years its
success was only limited to the surface and shallow interior. Surface
meridional flow, with a poleward speed of around $10 - 20$ \ms, was first
obtained from surface Doppler measurements \citep{Duv79, Hath96, Ulr10}. 
The result was confirmed by analysis through
tracking different surface features, such as small magnetic
elements \citep{Komm93, meu99}, sunspot groups \citep{how86, Woh01}, and
supergranules \citep{sva07, Hath12}. The meridional flow below the photosphere, 
 typically shallower than the depth of $\sim 30$ Mm,  was studied by       
use of helioseismic techniques, such as time--distance helioseismology 
\citep{Giles97, zha04}, Fourier-Legendre analysis \citep{Braun98, Roth16}, 
and ring-diagram analysis \citep{Gonz99, Har00, Bas03}. All these 
analyses gave consistent results that the meridional flow remains 
poleward in latitude up to about $50\degr$ from the surface to at least 30 Mm 
beneath the surface.

The efforts of inferring the meridional-circulation profile in the Sun's 
deeper interior are still ongoing. The first of such an attempt was made 
by \citet{Giles99} using data from the {\it Solar and Heliospheric 
Observatory} / Michelson Doppler Imager \citep[{\it SOHO}/MDI;][]{Scher95}, 
long before a systematic center-to-limb (CtoL) effect was found \citep{zha12}. 
Poleward flows were detected through nearly the entire convection zone, and 
the equatorward return flow, of 2 \ms, was only obtained near the base 
of the convection zone after applying in his inversions a constraint that the poleward 
flowing mass must be balanced by the equatorward flowing mass. A picture 
of single-cell circulation with a deep equatorward flow in each hemisphere
was thus suggested. However, this single-cell picture was challenged 
by some recent work. Through tracking supergranules using the MDI Doppler 
data, \citet{Hath12} reported a much shallower equatorward flow at a 
depth of $> 50$ Mm, just below the surface shear layer. However, in this 
method the depth was empirically determined by the supergranular anchoring depth, 
which is controversial and is also limited in the capability of reaching deeper areas,
hence helioseismic analysis is needed for a more robust inference of the 
meridional-circulation profile.

Recent progress in helioseismically inferring the Sun's deep meridional 
circulation was made after the systematic CtoL effect was found
and removed. Through analyzing the first 2 years of the {\it 
Solar Dynamics Observatory} / Helioseismic and Magnetic Imager 
\citep[{\it SDO}/HMI;][]{Scher12, Schou12} observations by use of the 
time--distance analysis technique, \citet{zha13} reported a detection of 
equatorward flow between 0.82 R$_\sun$ (a depth of $\sim 125$ Mm) and 
0.91 R$_\sun$ ($\sim 65$ Mm). Above and below this layer, the flows are 
mostly poleward, indicating a double-cell circulation in both hemispheres. 
Following a similar analysis procedure but using GONG observations, 
\citet{kho14} and \citet{Jac15} reported that the poleward flow turns to
equatorward at about the same depth as reported by \citet{zha13} but 
did not find a persistent poleward flow beneath the layer of equatorward flow. 
Later, through inverting time--distance measurements that are 
obtained from the HMI's 4 years continuous observations, with the radial 
flow component included and a mass-conservation condition applied, 
\citet{Raj15} reported that the equatorward flow was found only beneath 
0.77 R$_\sun$, consistent with a single-cell circulation. Despite the 
similarities in the procedures employed by the above-mentioned 
time--distance analyses, these authors reported results 
that are not fully consistent, highlighting the great challenges in 
inferring the Sun's meridional-circulation profile in the deep interior, 
as well as the necessity of a more reliable analysis strategy. 

Meanwhile, it is important to point out that all the above-mentioned 
time--distance analyses, despite their discrepancies in final profiles of 
the meridional circulation, positively 
reported a detection of equatorward flow, which is a significant progress 
after many years of fruitless searches. However, such a detection is not 
possible without removing the systematic CtoL effect in 
the measured acoustic travel times. The CtoL effect exhibits as extra 
travel-time shifts, in the same sense as what a poleward meridional 
flow would cause but much stronger \citep{zha12}. An empirical 
effect-removal method, i.e., using the east-west travel-time measurements 
along the Sun's equator as a proxy of the CtoL effect, was suggested 
\citep{zha12}, and most of the recent efforts studying the meridional 
circulation \citep{zha13, kho14, Jac15, Raj15, Chou_b} 
adopted this empirical method. However, how robust this empirical 
effect-removal method is, whether the CtoL effect is isotropic in all azimuthal 
directions to warrant such a removal, and what role the effect-removal process 
plays in causing the discrepancies in the results by different authors, 
are all questions remaining to be answered.

In addition to the time--distance analysis efforts introduced
above, a global helioseismology method using mode coupling 
was also developed to investigate the deep meridional circulation
\citep{sch11a, sch11b}. Their analysis on MDI data, 
covering the period of 2004 to 2010, revealed multiple circulation cells 
in both latitudinal and radial directions \citep{sch13}. Such analyses
provide valuable results that can cross examine the time--distance 
helioseismic results on the Sun's deep meridional-circulation profile.

In this paper, we introduce a new time--distance measurement strategy that
removes the CtoL effect more robustly than the previous empirical method of 
using a proxy. Exhaustively measuring acoustic travel times along all the solar 
disk's radial directions, we are able to set up linear equations that 
relate the measurements to the CtoL effect and the meridional-flow-induced 
travel-time shifts, both of which can be solved out from the linear 
equations. We then evaluate the isotropy of the CtoL effect, and invert 
the meridional-flow-induced travel-time shifts for the meridional-circulation 
profile throughout the convection zone. This paper is organized 
as follows: we introduce our new measurement method in \S2, and prepare 
data and perform measurements in \S3. In \S4, we disentangle the CtoL 
and the meridional-flow-induced travel-time shifts from raw measurements, 
and show our inverted meridional-flow profiles in \S5. In the end, 
we discuss our results in \S6.

\section{Measurement Method}
\label{sec2}

Time--distance helioseismology measures the time it takes acoustic 
waves (or $p$-mode waves) traveling from one surface location to another 
\citep{Duv93} after propogating through the Sun's interior. The travel-time shifts 
between two oppositely traveling waves with same surface ends are usually 
thought to be caused by interior flows, and are often used to invert
for the internal flow velocities. However, when inferring the Sun's meridional 
circulation, the travel-time shifts measured along the Sun's latitudinal 
direction, which were initially thought corresponding only to flows of the 
north-south direction, is actually a mixture of both meridional-flow 
induced time shifts and a systematic CtoL effect, and these two types of 
travel-time shifts need to be separated before they can be used for 
inversions \citep{zha12}. Here, we develop a new method to separate the 
CtoL effect from the meridional-flow-induced travel-time shifts through 
comprehensive measurements.

\begin{figure}[!t]
\epsscale{0.96}
\plotone{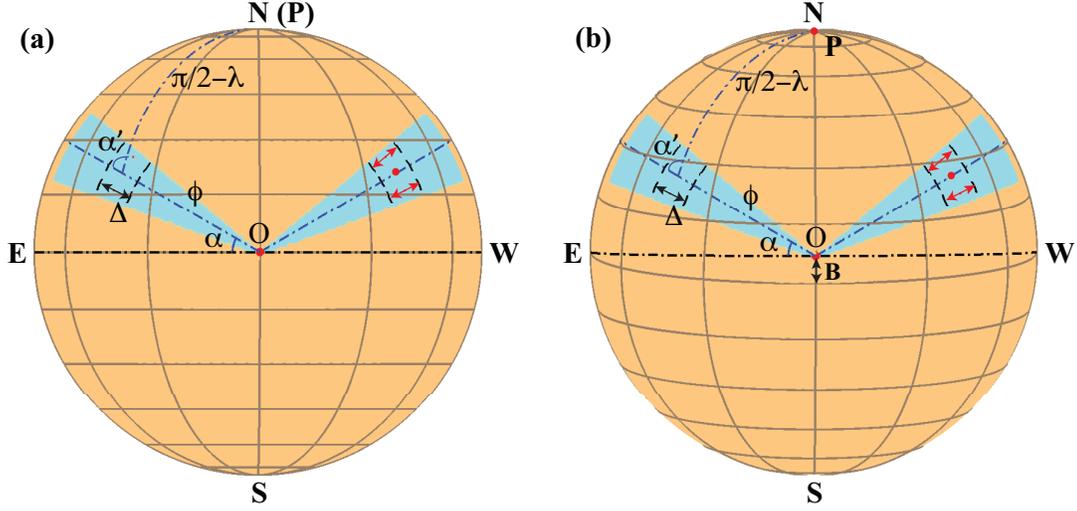}
\caption{Time--distance helioseismic measurement schemes for the cases of 
(a) zero solar B-angle and (b) non-zero solar B-angle. In each blue-shaded
area, we measure acoustic travel times between a pair of concentric arcs, 
which are at a distance of $\Delta$ apart and whose concentric center is 
at a distance of $\phi$ from the disk center. The line connecting the middle
points of both arcs lie along the radial direction of the solar disk, with an angle 
of $\alpha$ relative to the disk's apparent horizontal axis, and an angle of 
$\alpha^\prime$ relative to the local north at the concentric center.}
\label{scheme}
\end{figure}

Our new measurement strategy is to make exhaustive time--distance 
measurements for acoustic waves traveling between any two locations 
along solar disk's radial direction. The measured quantities are travel-time shifts,
$\delta\tau$, for a complete set of geometric parameters. As shown in 
the configuration in Figure~\ref{scheme}, the acoustic source and receiver, 
represented by a pair of concentric arcs lying along the disk's radial 
direction at an azimuthal angle $\alpha$ relative to the the disk's apparent 
horizontal axis, are a 
great-circle distance $\Delta$ apart. The concentric center of the arc pair 
is at a great-circle distance of $\phi$ from the disk center, and each arc spans 
$30\degr$ with a distance of $\Delta/2$ to the concentric center. To measure 
travel times for p-mode waves traveling between such a pair of arcs,
we cross-correlate the two time sequences averaged from the 
Doppler signals in the two arcs. From the resulting 
cross-correlation functions we fit for wave-traveling times for both 
traveling directions using the Gabor wavelet function. To enhance the 
stability of the Gabor wavelet fitting, we average all the cross-correlation 
functions over a wedge of a $20\degr$-wide azimuthal angle (see 
Figure~\ref{scheme}) in each Carrington rotation, about 27 days.

The differences between the fitted travel times in the opposite traveling 
directions, $\delta\tau$, in principle contain contributions from various 
causes, listed in a magnitude-decreasing order: rotation, CtoL effect, 
meridional flow, and maybe some other yet unknown factors. To eliminate 
the rotation, we average the symmetric time-shift measurements on 
either side of the central meridian, as shown in Figure~\ref{scheme}, 
since the projected components of the rotation along the two symmetric 
measurement directions are expected to have same values but opposite 
signs. The symmetrized time shift $\delta\tau(\alpha, \phi, \Delta)$ 
becomes rotation-free, and it can be related to the CtoL effect, \dtc, 
and the meridional-flow-induced travel-time shifts (as measured along 
latitudinal direction), \dtm, by a linear equation,
\begin{equation}
\delta\tau(\alpha,\phi,\Delta) = \delta\tau_\mathrm{CtoL}(\phi, \Delta)
         + \delta\tau_\mathrm{MF}(\lambda, \Delta) \, \cos\alpha^\prime,
\label{le}
\end{equation}
where $\lambda$ is latitude of the cocentric center of the arc pair, and 
$\alpha^\prime$ is the angle between the measurement direction (disk 
radial) and the local north direction (see Figure~\ref{scheme} 
and Equation~\ref{geo2}). In this equation, the \dtc \, is assumed 
azimuthally symmetric, i.e., invariant to azimuthal angle 
and varies only with $\phi$ and $\Delta$, and its validity will be 
examined later in \S\ref{sec4}. The \dtm \, is 
longitudinal symmetric and varies with $\lambda$ and $\Delta$. When 
measured at an angle $\alpha^\prime$ relative to the local north, only 
the projection of meridional flow along the ray path contributes (under 
ray-path approximation as discussed in \S\ref{sec4}), i.e., the time-shift 
contribution to the measured time shift is $\delta\tau_\mathrm{MF} \, 
\cos\alpha^\prime$. On the observed disk with a solar B-angle $B$, the 
$\lambda$ and $\alpha^\prime$ in Equation~\ref{le} are given by a spherical 
geometric relation: 
\begin{equation}
\lambda=\sin^{-1}{[\cos{\phi}\sin{B}+\sin{\phi}\cos{B}
\sin{\alpha}]}
\label{geo1}
\end{equation}
\begin{equation}
\alpha^\prime=\cos^{-1}{\bigg{[}\frac{\cos{\phi}\sin{\lambda}-\sin{B}}
{\sin{\phi}\cos{\lambda}}\bigg{]}}.
\label{geo2}
\end{equation}

Our processed data images are in the Postel's projection coordinates, 
with the observed disk center
as the coordinate origin, and the equator is off the apparent horizontal 
direction when the B-angle is non-zero. The CtoL effect depends on the 
location on the solar disk and is thus irrelevant to the solar B-angle, 
while the meridional flow is directly dependent on latitude that varies 
in the disk location with the  B-angle. Therefore, the measurements at 
the same disk location over time cannot be averaged directly. However, 
since the B-angle varies slowly, we average the measurements for each 
Carrington rotation without expecting significant side effects, and the 
mean B-angle for each rotation is used when solving the equations above.

\section{Data and Measurements}
\label{sec3} 

\subsection{Data Preparation}
\label{sec31} 

\begin{figure}[!t]
\epsscale{0.90}
\plotone{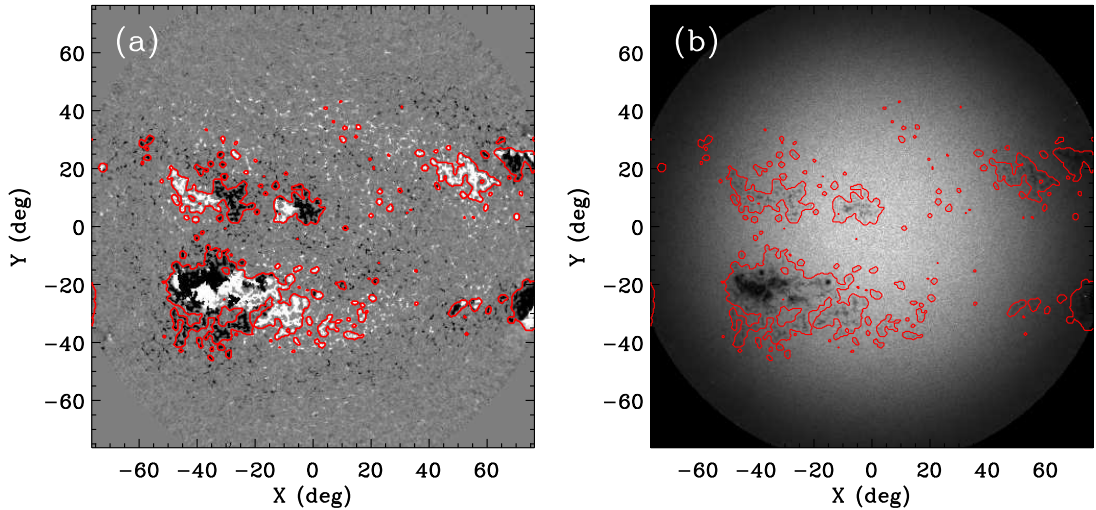}
\caption{(a) Line-of-sight magnetic field image taken on 2012 July 8,
with green contours of 10 G, inside which Doppler data are not used 
in our time--distance measurements. (b) Acoustic power map for the same 
date with the same contours over-plotted.}
\label{mask}
\end{figure}

We use 7 years of HMI full-disk Doppler-velocity data, from 2010 May 1
to 2017 April 30, for this analysis. The data are organized into 24-hr chunks
with a 45-s cadence. The original data are first rebinned to $1024 \times
1024$ pixels, and then tracked with the Carrington rotation rate and 
projected into Postel's projection coordinates, with projected image center being
the observed disk center. The tracked and remapped data are then rebinned 
again to a spatial resolution of $0\fdg36$ pixel$^{-1}$ (here, 
$1\degr$ is 1 heliographic degree) at the disk center in order to reduce the 
computational burden in the following time--distance computations while not losing 
resolution for the deep interior detections. We then get running-difference 
images, the difference between one image and the image immediately ahead 
of it, to reduce the effect of solar convection, so that the data can 
be used for helioseismic analysis without further filtering in the Fourier 
domain. An acoustic power map from one random date is shown in 
Figure~\ref{mask}b.

Another effect one needs to consider is the surface magnetic effect in 
helioseismic measurements, as recently demonstrated by \citet{Dingyi_M}. 
The measured travel times for acoustic waves traveling into and out 
from a magnetic region exhibit a significant asymmetry that is, presumably,
not accounted for by large-scale horizontal flows. This non-flow-related 
time shift will bias our meridional-flow-induced time-shift measurement 
notably if one end of the wave path is located inside a magnetic region. Therefore 
the magnetic regions should be masked to avoid such activity-related 
artifacts, as suggested by \citet{Dingyi_M}. To prepare the masks, we 
track and remap the line-of-sight magnetic field data in the same way 
as for the Doppler-velocity data, but with a lower time step of 2 hours, 
and then average the daily images and smooth them using a normalized 2D 
Gaussian function kernel with a FWHM  of $1\fdg8$. Data falling into 
the area where magnetic field strength exceeds a threshold of 10~G will 
be masked and not used in the following helioseismic analysis.
At most $14.5\%$ of the data pixels are masked during the solar 
maximum and a negligible number of pixels during the minimum years.
The masking reduces the amount of 
data and boosts the noise level, therefore cannot be applied aggressively. 
The threshold of 10~G is so chosen that the deficit in acoustic power 
in and around the active regions are mostly covered, as shown in 
Figure~\ref{mask}.

\subsection{Time--Distance Helioseismic Measurements}
\label{sec32}

\begin{figure}[!t]
\epsscale{1.00}
\plotone{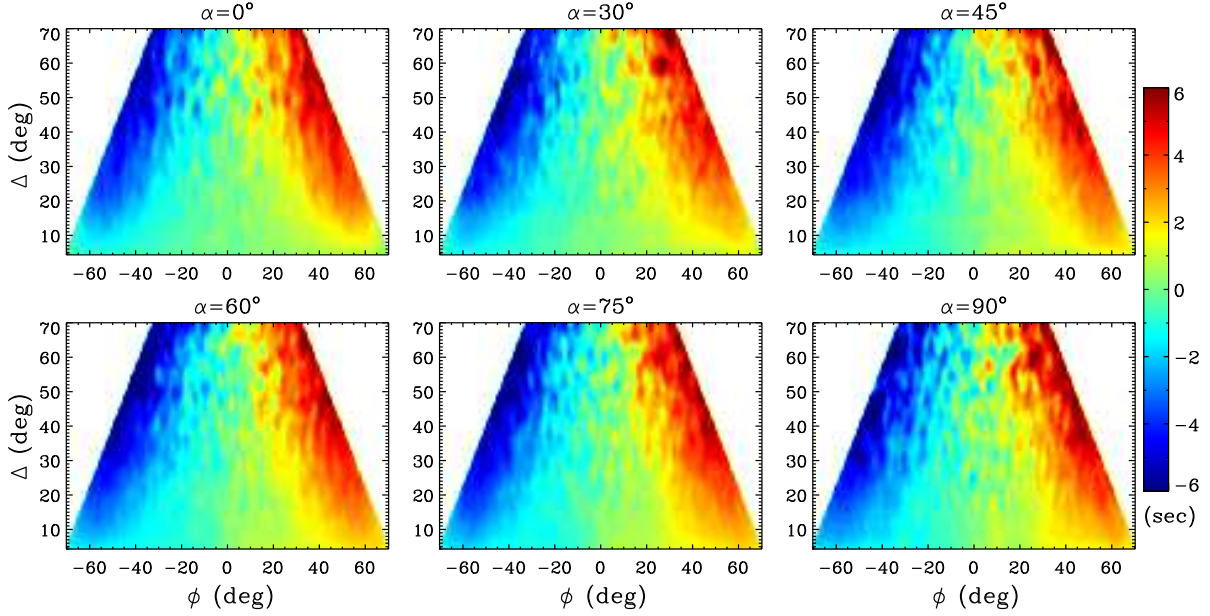}
\caption{Seven-year averages of $\delta\tau$ measurements for each azimuthal
angle $\alpha$ that is used in our analysis.}
\label{raw}
\end{figure}

Following the measurement procedure prescribed in \S\ref{sec2}, one set of $\delta 
\tau(\alpha, \phi, \Delta)$ is obtained for each Carrington rotation 
starting from 2010 May 1, and a total of 93 rotations are obtained for 
the 7-yr period. Figure~\ref{raw} shows the 7-yr-averaged $\delta\tau(\phi, \Delta)$  
for the 6 azimuthal angle $\alpha$'s that are used in our following analysis. 
Since the east-west symmetric measurements (of $\alpha$ and $180\degr-\alpha$) 
are folded to cancel the rotation, the final $\alpha$ only ranges from 
$0\degr$ (east) to $90\degr$ (north). All measurements in certain azimuthal wedges 
are binned into $\alpha = 0\degr, 30\degr, 45\degr, 60\degr, 75\degr$, 
and $90\degr$, with $\alpha=15\degr$ discarded due to that this azimuthal wedge 
mostly lie in the activity belts for many analysis periods and the measurement noises 
are much higher than for other $\alpha$'s.

For each panel of Figure~\ref{raw}, the horizontal axis is the disk-centric 
distance $\phi$, ranging from $-65\degr$ to $65\degr$ with a distance 
sampling of $1\fdg08$, and the vertical axis is skip distance $\Delta$, 
ranging from $4\fdg32$ up to about $70\degr$ with a step of $0\fdg72$. The 
$\phi$ has positive sign in the northern hemisphere, and it is same 
as longitude in the case of $\alpha=0\degr$, and same as latitude  
in the case of $\alpha=90\degr$. The measurements have values in a 
trapezoid area because for larger measurement distances, one end of 
the measurement pair is too close to or out of the solar limb.
The $\delta\tau$ is the difference between the two opposite traveling 
directions, with waves traveling from larger $\phi$ to smaller $\phi$ 
as positive direction.

All the panels in Figure~\ref{raw} show a general anti-symmetric pattern 
with positive $\delta\tau$ for positive $\phi$ and vice versa. These 
panels are dominated by the CtoL effect, which exhibits as longer 
travel times along the direction from limb to disk center than along the opposite
 direction. The variations among different panels, most noticeable 
visually for small skip distances, are caused by the \dtm\, projected into the 
disk's radial direction. For $\alpha=0\degr$, there is little contribution from 
the meridional flow to the measured  $\delta\tau$, and the measurements 
were empirically used as the CtoL effect in the previous studies 
\citep[e.g.,][]{zha13}, while for $\alpha=90\degr$, $\delta\tau$ is a 
combination of the \dtc\, and the \dtm. If only using measurements 
with $\alpha=0\degr$ and $\alpha=90\degr$, our method 
essentially reduces to the method of those previous time--distance 
studies, but with a couple of differences. One difference is 
that our measurements are always along the disk's radial direction, 
while the previous authors measured along the north-south directions, 
which are off the radial direction except on the central meridian.  The other 
difference is, the previous methods applied an averaging within a wide 
stripe, and the stripe was $30\degr$-wide in 
longitude on the equator and $60\degr$-wide in longitude at $65\degr$ 
latitude. Our averaging is in a narrow $20\degr$-wide azimuthal wedge, 
which is 1 pixel wide at the disk center and about $20\degr$-wide in 
longitude at high-latitude regions. This narrower averaging pattern 
is designed for less overlapping in multiple-$\alpha$ measurements. 
However, we also realize that as a trade-off, our single-$\alpha$ 
measurement has larger noises for each individual wedge. Overall, 
our exhaustive measurements capture more information than the 
previous method on the intertwine of 
the \dtm\, and \dtc\, that the observations provide.

Due to the solar B-angle variation, throughout the 93 Carrington rotations, 
each pixel in the panels is measured at the same disk location $(\alpha, 
\phi)$ but not a same latitude. The maximum variation in latitude for one 
certain disk location can be $\pm7.2\degr$, and this effect is not negligible. 
Therefore, on the 7-yr-averaged images, shown in Figure~\ref{raw}, 
\dtc\, are exactly averaged while \dtm\, are smeared. Here we 
average these raw measurements to show the general patterns, but in 
the next section  we solve a set of linear equations, 
given in Equation~\ref{le} -- \ref{geo2}, with varying B-angles, to infer 
the $\phi$-dependent \dtc\, and the latitude-dependent \dtm.

\section{Disentangling CtoL Effect and Meridional-Flow-Induced Travel-Time
Shifts}
\label{sec4} 

\begin{figure}[!t]
\epsscale{0.95}
\plotone{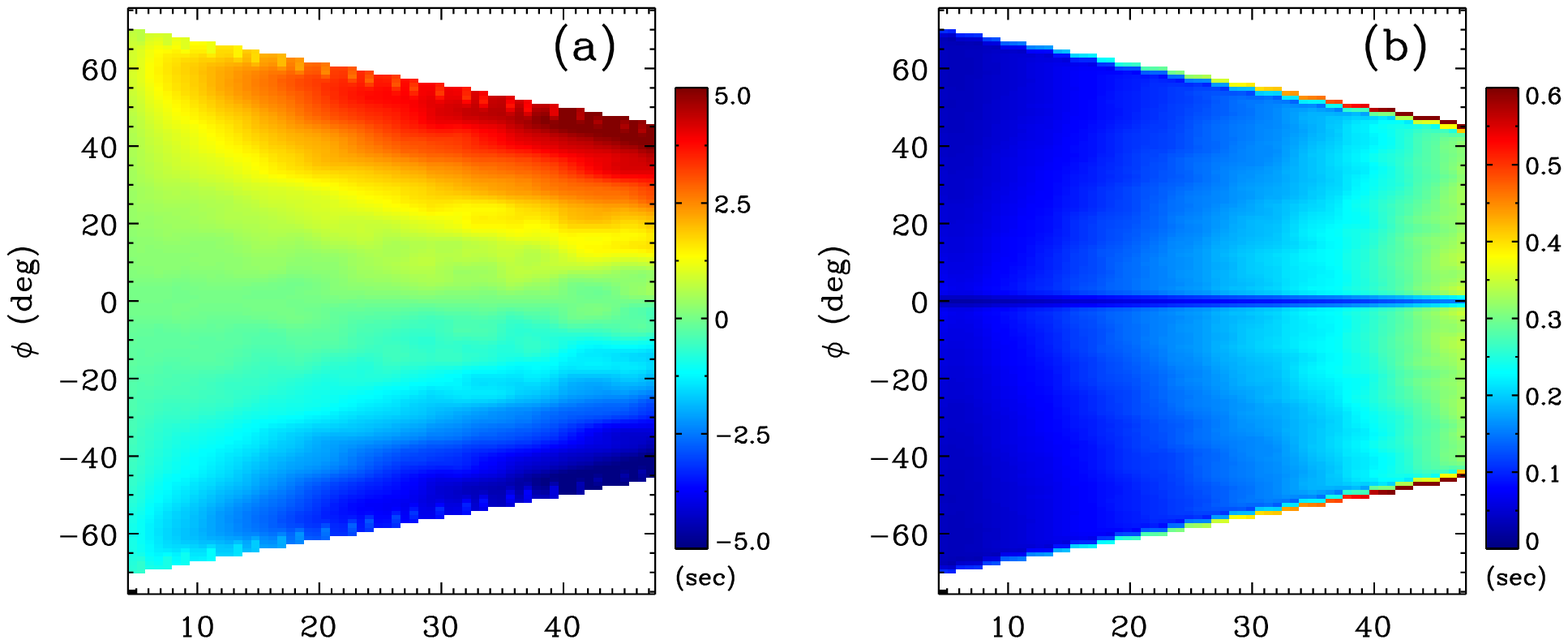}
\plotone{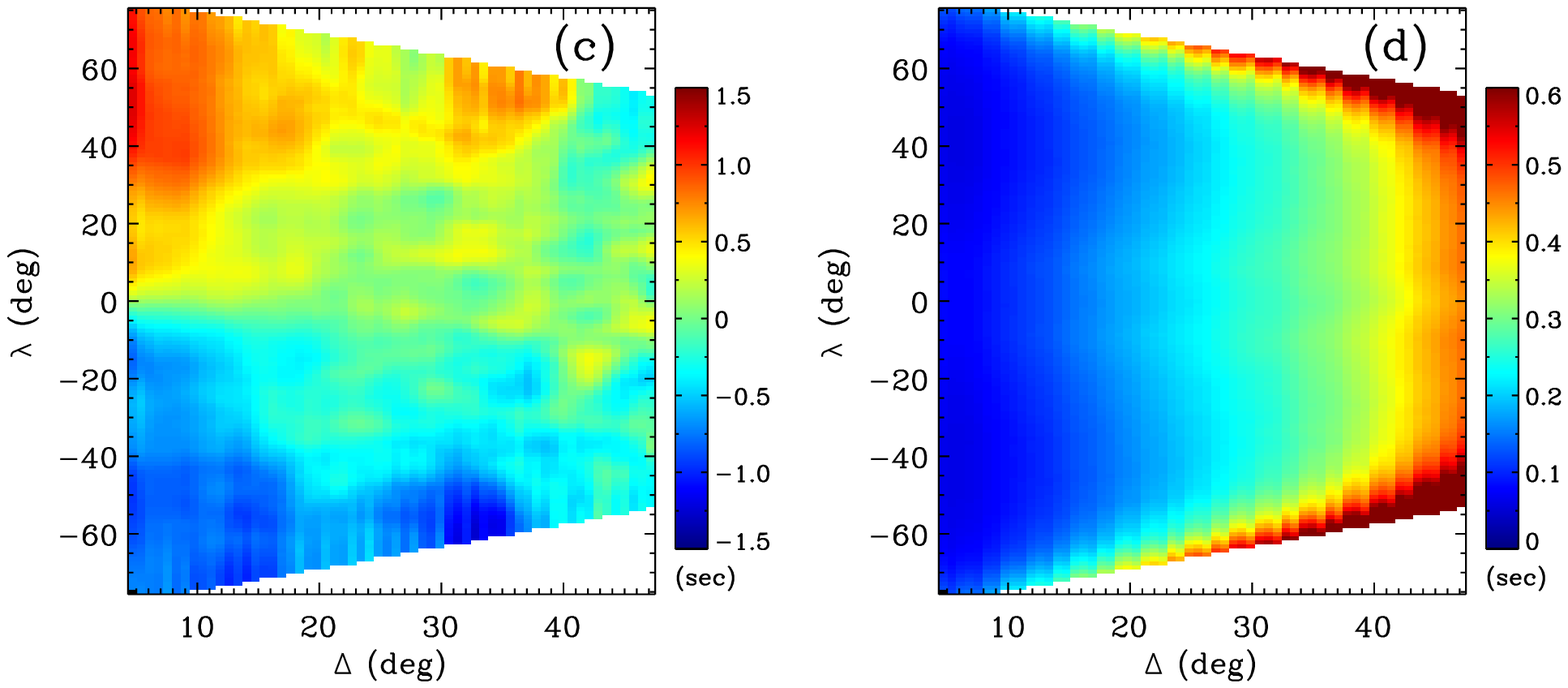}
\caption{(a) CtoL effect \dtc\ disentangled from our measurements. (b) Uncertainties
of the \dtc. (c) Meridional-flow-induced travel-time shifts 
\dtm\, disentangled from our measurements. (d) Uncertainties of the \dtm. }
\label{CtoLfl}
\end{figure}

The \dtc\, and \dtm\, are disentangled by solving linear equations 
Equation~\ref{le} -- \ref{geo2}. We observe that Equation~\ref{le} holds 
independently for each $\Delta$, that is, the subgroup of 
$\delta\tau(\alpha,\phi,\Delta)$ measured with a specific $\Delta$ 
forms a subset of linear equations, in which the unknown quantities 
\dtc\, and \dtm\, are both one-dimensional varying only with 
$\phi$ and $\lambda$, respectively. Besides, the combination of $\alpha$ 
and $\phi$ in $\delta\tau$ covers a wide range of $\phi$ and $\lambda$, 
and $\phi$ and $\lambda$ vary separately, making it an 
over-determined linear problem. We solve this over-determined one-dimensional 
linear equation set by employing a standard least-square method with 
a 2nd-order Tikhonov regularization on \dtc\, and a 1st-order Tikhonov 
regularization on \dtm\ \citep{Ast05}. 
We assume \dtc\, to be azimuthally symmetric, thus \dtc$=0$ at the disk center, 
and these prior information is incorporated as a constraint in the 
regularization as well.

\begin{figure}[!t]
\epsscale{0.60}
\plotone{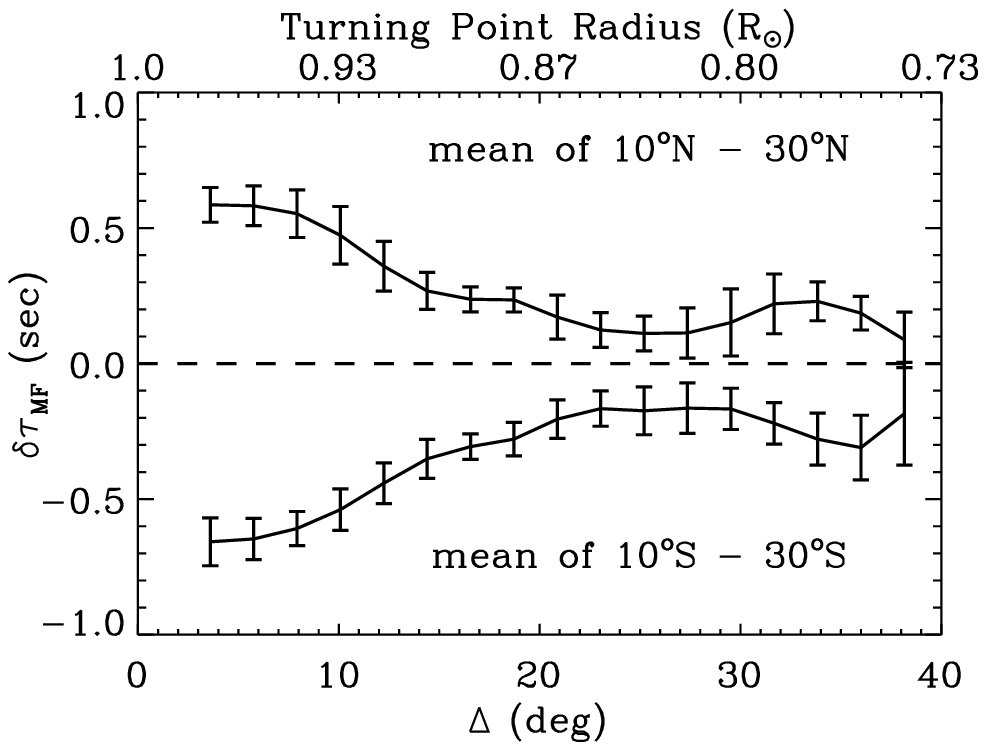}
\caption{ Average curves of $\delta\tau_\mathrm{MF}$ for $10\degr - 30\degr$
latitudinal bands in both hemispheres. }
\label{dt_fl}
\end{figure}

Figure~\ref{CtoLfl} shows the $\delta\tau_\mathrm{CtoL}(\Delta, \phi)$
and $\delta\tau_\mathrm{MF}(\Delta, \lambda)$, which are disentangled from
our travel-time measurements through solving the linear equation set for 
the 7-yr period, and their uncertainties. The \dtc,
of an order of 5~s, increases with skip distance $\Delta$ and disk-centric 
distance $\phi$, and its uncertainty increases with $\Delta$ as well. 
In each hemisphere, the \dtm, of an order of 1~s, 
shows a trend of decrease in magnitude with $\Delta$ within the range 
of $0\degr$ to about $20\degr$. Then the trend reverses to increase till 
the skip distance of about $35\degr$. To show this trend more clearly, 
we plot the average curves of \dtm, for latitudinal 
bands of $10\degr - 30\degr$ in both hemispheres (see Figure~\ref{dt_fl}). 
The decreasing trend of magnitudes with skip distances smaller than 
$25\degr$ indicates that the poleward flow becomes weaker or possibly 
turns equatorward at shallow depths. That the \dtm\,
curves bend back beyond a skip distance of $30\degr$ indicates that the 
flow velocity reverses to poleward again below about 0.80 R$_{\sun}$.

As can be seen in Figure~\ref{CtoLfl}, at large skip distances the 
\dtm\, is one order of magnitude smaller than \dtc, and the uncertainty 
of \dtm\, increases greatly for large skip distances. Therefore the \dtm\,
beyond $\Delta=40\degr$ are not used in our meridional-flow 
inversions in \S5. We do not assume a hemispheric symmetry for the 
\dtm\, when solving Equation~\ref{le}, and the solved \dtm\, indeed 
shows a north-south asymmetry, as can be seen in Figure~\ref{CtoLfl}c. 

\begin{figure}[!t]
\epsscale{0.98}
\plottwo{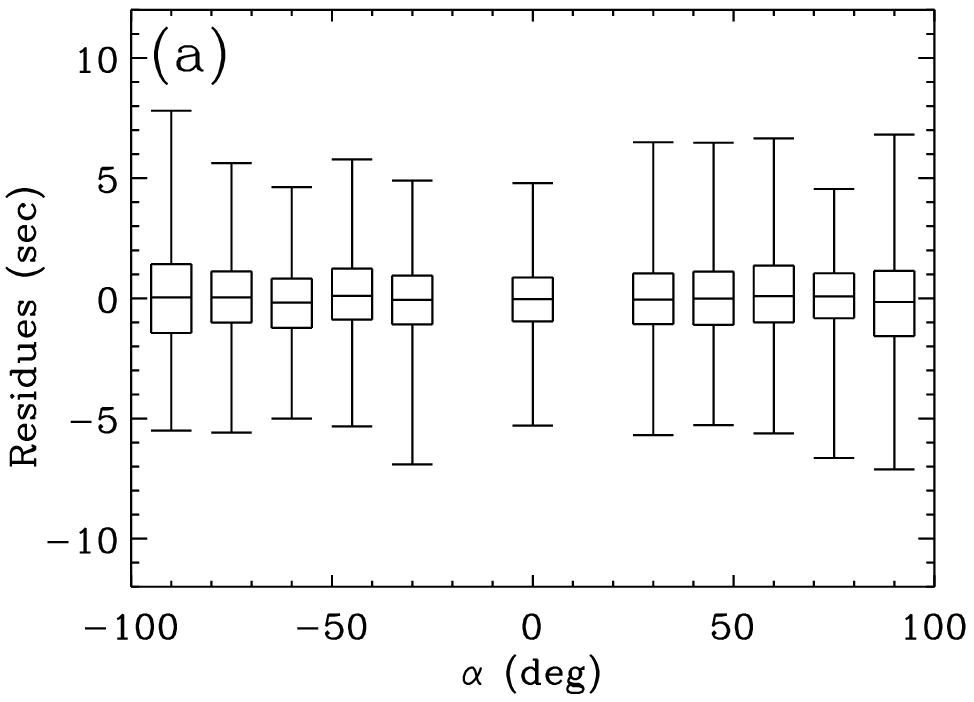}{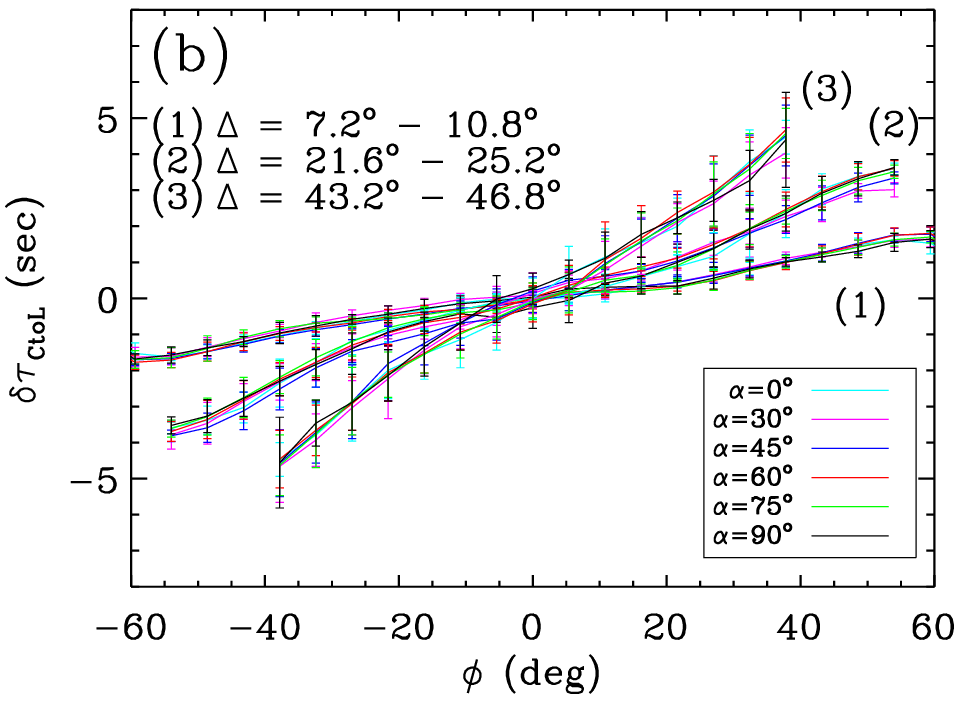}
\caption{Left: Distribution of residues in model fitting, for $\phi=30\degr$
and $\Delta$ falling between $14\fdg4 - 21\fdg6$. Right: CtoL effect
derived for different $\alpha$'s for 3 skip-distance ranges. }
\label{CtoLsym}
\end{figure}

In Equation~\ref{le} -- \ref{geo2} the \dtc\ is assumed azimuthally symmetric, 
and here we examine the validity of this assumption. 
If this assumption does not hold and \dtc\ varies with $\alpha$, then Equation~\ref{le}
cannot fit travel--time measurements well and the misfit shall show a dependence 
on $\alpha$. We examine the residues of fitting, $\delta \tau(\alpha, \phi, \Delta) 
-(\delta\tau_\mathrm{CtoL}(\phi, \Delta)
+ \delta\tau_\mathrm{MF}(\lambda, \Delta) \, \cos\alpha^\prime)$, where 
$\delta\tau_\mathrm{CtoL}$ and $\delta\tau_\mathrm{MF}$ are
solved from the linear regression. Figure~\ref{CtoLsym}a shows 
the distribution of the residues over different $\alpha$'s for 
$\phi=30\degr$ and $\Delta$ between $14\fdg4 - 21\fdg6$. The box-plot 
shows the sample minimum, lower quartile, median, upper quartile, and 
maximum of the residues. The residues are randomly distributed 
and do not show any tendency with $\alpha$, indicating 
that the assumption of an isotropic CtoL is reasonable, i.e., the CtoL 
effect in the time--distance measurements using the HMI Doppler 
observations is azimuthally uniform within our measurement uncertainties.
As another way to check, we examine the $\alpha$-dependent 
\dtc, which are obtained by subtracting the 
meridional-flow contribution $\delta\tau_\mathrm{MF} \cos\alpha'$ (after 
$\delta\tau_\mathrm{MF}$ is solved) from the raw measurements of 
$\delta\tau$ for each $\alpha$. As shown in Figure~\ref{CtoLsym}b, for 
3 selected skip distances, the CtoL effects agree well for different 
azimuthal angles.

\section{Inverting for Meridional Flow}
\label{sec5} 

Meridional flow can be inverted from solving linear equations that relate 
the travel-time shifts $\delta\tau$, measured between two surface locations 
$\mathbf{x_1}$ and $\mathbf{x_2}$, and the flow velocity field $\mathbf{v}
(\mathbf{r})$:
\begin{equation}
\delta\tau({\mathbf{x_1}, \mathbf{x_2}}) = 2 \int K(\mathbf{r}, \Delta)  
\, \mathbf{v} (\mathbf{r^\prime}-\mathbf{r}) \, \mathrm{d}\mathbf{r},
\label{Ker_equ}
\end{equation}
where $K(\mathbf{r}, \Delta)$ is a three-dimensional 
flow sensitivity kernel for travel distance $\Delta$, which quantifies the 
amount of travel-time shifts caused by a localized unit flow, 
$\Delta = |\mathbf{x_1} - \mathbf{x_2}|$ is the great-circle 
distance between two surface locations $\mathbf{x_1}$ and $\mathbf{x_2}$,   
$\mathbf{r'}=\frac{1}{2} (\mathbf{x_1}+\mathbf{x_2})$
is the midpoint of $\mathbf{x_1}$ and $\mathbf{x_2}$, and 
$\mathbf{r}$ is the offset between $\mathbf{r'}$ and the flow location.
Since the kernel $K(\mathbf{r}, \Delta)$ is spherically 
symmetric for the same great-circle distance $\Delta$, it is denoted as 
$K_\Delta(\mathbf{r})$ hereafter. For $K_\Delta(\mathbf{r})$, 
we adopt the widely-used ray-path 
approximation kernels \citep[e.g.,][]{Ray_path, zha13}, in which the 
wavelength is assumed negligible relative to the flow length scale and 
the sensitivities concentrate only on ray paths. Ray-path approximations 
clearly have its limitation, and wave-based kernels, such as recently 
developed by \citet{Bon16} and \citet{Giz17}, will be used to replace the 
ray-path kernels in our future inversions. 

In this paper, following practices by some previous authors \citep{zha13,
Jac15}, we do not include radial flow $v_r$ in our inversions for the 
following reasons. First, the radial-flow-induced signals are expected 
to be at least one order of magnitude smaller than those caused by the 
meridional flow. Second, the travel-time shifts are only sensitive to the 
differences in radial flows, i.e., a uniform radial flow give zero time 
shifts since the effects along the upward and downward portions of the 
ray path get canceled. Third, cross talk between horizontal and radial 
flows, e.g., a downward flow and poleward flow in shallow interior for 
a poleward traveling wave play a same role in increasing \dtm, 
is another factor that complicates the inference of the radial flow.

Considering only the horizontal meridional flow $v_\theta$, 
Equation~\ref{Ker_equ} can be rewritten as:
\begin{equation}
\delta\tau_\mathrm{MF} (\lambda, \Delta) =2 \int 
K_{\Delta}(l, z) \, v_\theta(\lambda - l, z) \, \mathrm{d}l\mathrm{d}z,
\label{Ker_rew}
\end{equation}
where $z$ is depth and $l$ is latitudinal offset relative to the latitude 
$\lambda$. Under ray-path approximation, the sensitivities concentrate only 
on ray paths, so the three-dimensional kernel $K_\Delta(\mathbf{r})$ in 
Equation~\ref{Ker_equ} reduces to a two-dimensional kernel $K_{\Delta}(l,z)$ 
in the meridional plane in Equation~\ref{Ker_rew}. Equation~\ref{Ker_rew} 
is actually a convolution that can be more easily solved in the Fourier 
domain than in the space domain. In the Fourier domain, the equation 
can be written as:
\begin{equation}
\widehat{\delta \tau}_\mathrm{MF} (k, \Delta) = 2 \sum_z \widehat{v}
(k, z) \, \widehat{K}_{\Delta} (k, z),
\label{Ker_rew2}
\end{equation}
where \, $\widehat{}$ \, denotes Fourier transform, and $k$ is spatial 
frequency in the latitudinal direction. It is clear that the equations 
become decoupled with $k$, and we get a one-dimensional 
linear equation for each $k$:
\begin{equation}
\widehat{\delta \tau}_\mathrm{MF} (\Delta) = 2 \sum_z \widehat{v} (z) 
\, \widehat{K}_{\Delta} (z).
\label{eq_fourier}
\end{equation}
We then solve each linear equation separately using a standard least-square
solver with a 0th-order Tikhonov regularization \citep{Ast05}. Since the radial 
flow is neglected in our equations, one should not expect the meridional flow $v$ 
obtained above to satisfy the local mass conservation. Instead, we consider a 
global mass conservation and limit the integrated mass flow on a
latitudinal cross-section to be close to 0, similar to what \citet{Giles99} did. 
This is used as a constraint in our regularization as well. 

\begin{figure}[!t]
\centering
\includegraphics[width=0.4\textwidth]{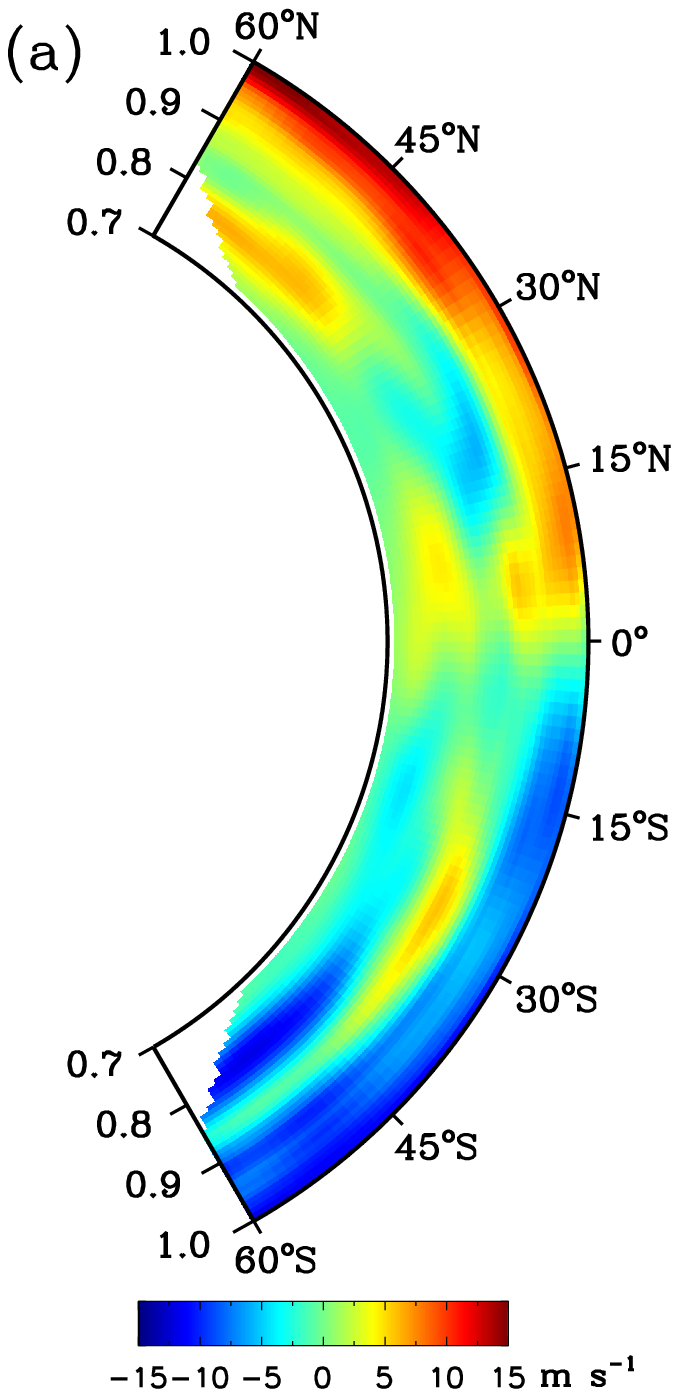}
\includegraphics[width=0.5\textwidth]{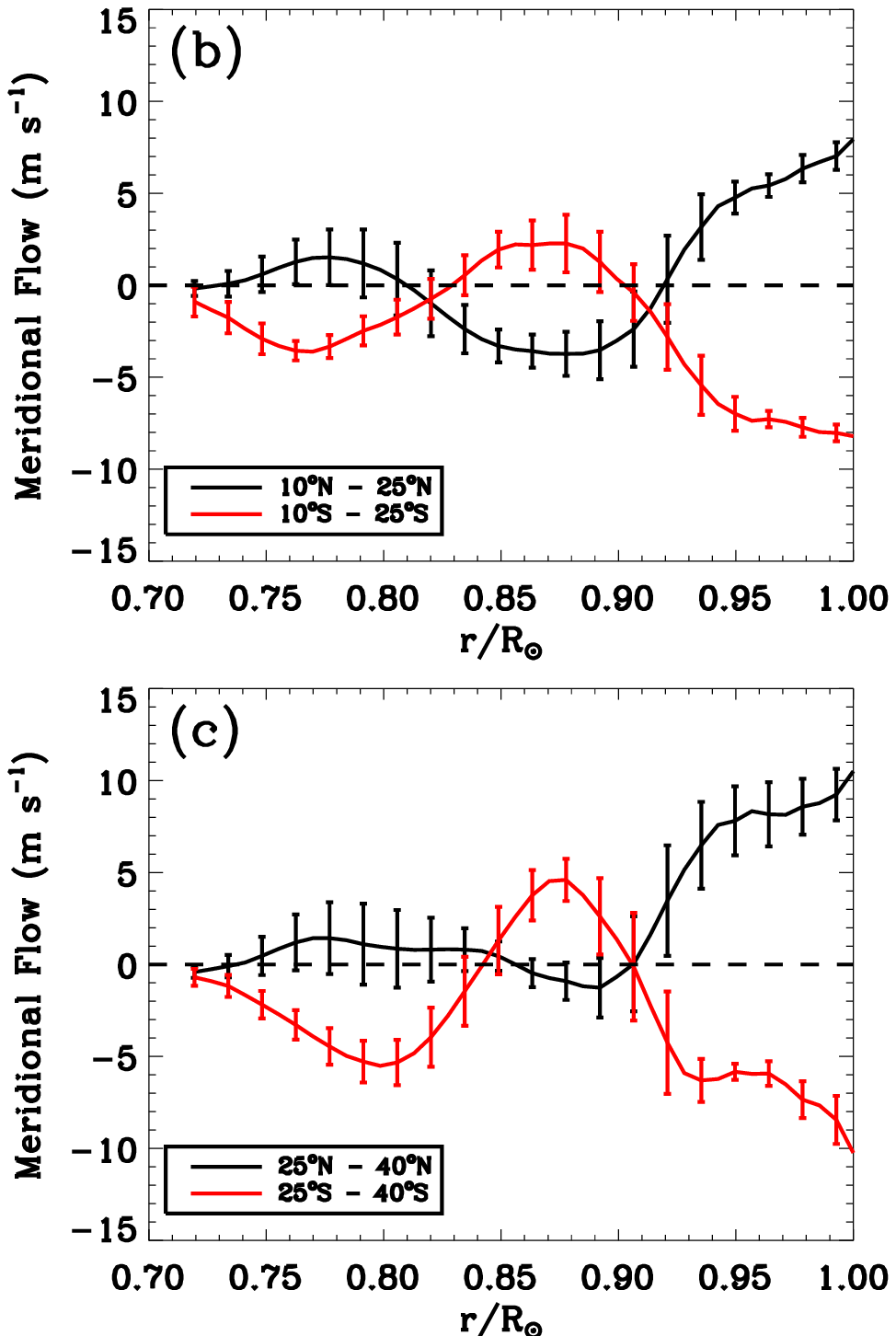}
\caption{(a) Inverted meridional flow in the convection zone, with positive flow
toward north. (b) Meridional-flow profile averaged from the $10\degr - 
25\degr$ latitudinal bands in both hemispheres. (c) Same as panel (b), but 
for the $25\degr - 40\degr$ bands.}
\label{Inv}
\end{figure}

Figure~\ref{Inv} shows our inversion results. In both hemispheres for all 
latitudes, the poleward flow, of about $10$ \ms, extends from the surface to 
about 0.91 R$_\sun$. Below 0.91 R$_\sun$, the flow turns equatorward with a 
speed of about $5$ \ms, extending to about 0.82
R$_\sun$ for low latitude regions, but not as deep for higher latitude. 
Beneath the layer of equatorward flow, the flow turns poleward again 
with a speed of lower than $5$ \ms. The averaged velocity profiles for 10$\degr - 25\degr$ 
and 25$\degr - 40\degr$ latitudinal bands, as shown in Figure~\ref{Inv}b 
and \ref{Inv}c respectively, better illustrate the two direction reversals 
at depths of about 0.91 R$_\sun$ and 0.82 R$_\sun$ for 10$\degr - 25\degr$ 
latitude, and 0.91 R$_\sun$ and 0.85 R$_\sun$ for 25$\degr - 40\degr$ 
latitude. The northern and southern hemispheres exhibit an asymmetry 
in these inverted flow results. Compared with the southern hemisphere, 
the equatorward flow in the northern hemisphere extends into deeper 
convection zone, but does not span as wide in latitude.

\begin{figure}[!t]
\epsscale{0.6}
\plotone{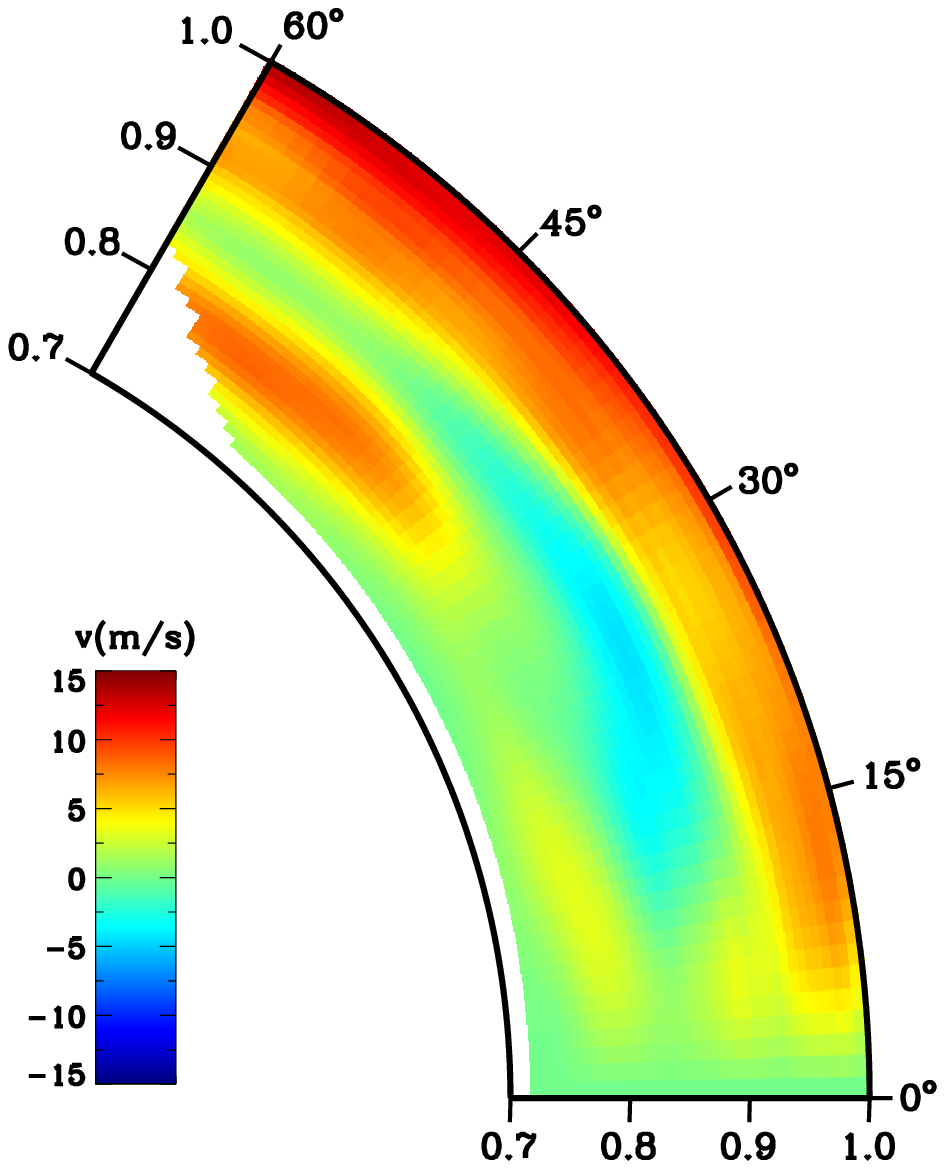}
\caption{Hemispherically symmetrized meridional-flow profile. Positive flow
is poleward and negative is equatorward.}
\label{Inv_sym}
\end{figure}

To more clearly display the general pattern of the meridional circulation,
we symmetrize the meridional-flow profiles of both hemispheres. The symmetrized
profile (Figure~\ref{Inv_sym}) shows a 3-layer flow structure,
indicating two circulation cells stacked in the radial direction in 
the convection zone. The surface poleward layer and the equatorward layer 
form the first circulation cell, and the lower part of the equatorward 
layer together with the deep poleward layer form the second cell, giving a
picture consistent with the cartoon plot Figure 1 by \citet{zha13}.

\section{Discussions} 
\label{sec6}

We have presented a new time--distance helioseismic measurement method 
to separate the systematic CtoL effect \dtc\, and the meridional-flow-induced 
travel-time shifts \dtm. The new method measures exhaustively acoustic 
travel-time shifts along the solar disk's all radial directions at all 
possible locations for all possible skip distances, and the \dtc\, and 
\dtm\, are disentangled from these measurements through solving 
a set of linear equations that relate these quantities to the measurements. 
The \dtm\ are then used to invert for the Sun's meridional circulation.
Applying the new method on the 7-yr HMI Doppler observations, we are able 
to get a more robustly determined CtoL effect and a more accurately 
measured \dtm. The meridional circulation, inverted from the \dtm, exhibits
essentially three flow layers that indicate a double-cell circulation 
in each hemisphere.

Our new measurement strategy is designed to give a more robust and accurate 
determination in the CtoL effect and thus the meridional flow than the previous method 
introduced in \S\ref{sec1}. As demonstrated in \S\ref{sec4}, the 
CtoL effect is isotropic relative to azimuthal angles, and this confirms 
the validity of the previous method -- removing a CtoL effect proxy that 
is assessed using time-shift measurements along the equator. However, as 
we know, the final inference of the meridional flow relies strongly upon 
the accuracy of the effect removal, and a small error in the CtoL effect 
can cause a big error in the inverted flow. The CtoL effect obtained by 
our new method, disentangled from the exhaustive measurements throughout the 
entire solar disk along all radial directions, is more robust than the 
proxy in the previous method, which measures the CtoL effect in a narrow band 
in the equatorial area.

It is noticeable that our results on the meridional-circulation profile 
do not fully agree with any of the published results introduced in 
\S\ref{sec1}, but more similar to the result by \citet{zha13} in the 
3-layer flow structure, although our flow speed in the deepest layer 
is substantially slower than theirs. One possible 
reason that these two sets of results have more similarities than others is 
that, the analysis period by \citet{zha13} was mostly during the 
solar minimum years, and magnetic field does not influence much on their 
final results,  while our analysis largely remove the effects
of the surface magnetic field following a method suggested by 
\citet{Dingyi_M}. In this regard, the periods analyzed by both 
\citet{Jac15} and \citet{Raj15} have stronger magnetic activities that may
complicate their final results. Meanwhile, it is not deniable 
that the major differences between our result and that of \citet{Raj15}
are likely caused by the inclusion of the radial flow and the mass-conservation
condition in their inversion.

Some improvements can be made in our future efforts of inverting the 
\dtm\, for the meridional circulation. One improvement is to include local-scale 
mass-conservation equations, for which the radial flow component is desired. 
We have reasonably neglected the radial component in this paper, but it is 
worth studying after a higher signal-to-noise ratio of the travel-time shifts is achieved in 
measurements and a method to disentangle the cross-talk between 
the horizontal and radial components is developed. 
Another recent development is the availability
of the wave-based sensitivity kernels for the meridional flow in spherical 
coordinates. \citet{Bon16} developed Born-approximation kernels and 
\citet{Giz17} developed wave-based kernels using numerical 
method, both of which are believed superior to the ray-approximation 
kernels that were used in the previous and current time--distance 
helioseismic inversions.  These wave-based sensitivity kernels will be 
used in our future efforts.

Moreover, in this paper, only the 7-yr-averaged results are inverted and 
shown. However, as the Sun experienced from the activity minimum to the 
maximum and back toward the minimum again during these 7 years, the 
meridional circulation may have also experienced detectable changes. 
Armed with our new analysis method, we are capable of inferring the 
meridional flow for shorter periods and analyzing the temporal evolution 
of the Sun's deep meridional circulation. That will offer us more 
knowledge of the Sun's internal dynamics for a better understanding of 
the solar dynamo and its cycles.

\acknowledgments {\it SDO} is a NASA mission, and HMI project is supported 
by NASA contract NAS5-02139 to Stanford University. R.C. is partly supported by 
the NASA Earth and Space Science Fellowship NNX15AT08H, and J.Z. is 
partly supported by NASA Grant NNX15AL64G.

\end{document}